\documentclass[twocolumn,superscriptaddress]{revtex4-2} 

\usepackage{graphicx} 
\usepackage{natbib} 
\usepackage[usenames,dvipsnames]{color} 
\usepackage{amsmath} 
\usepackage{amssymb} 
\usepackage{soul} 
\usepackage{ifthen} 
\usepackage[ampersand]{easylist}
\usepackage{braket}
\usepackage{todonotes}

\newcommand{\trace}[2]{\t{Tr}_{#1}\left(#2\right)}

\newcommand{\sch}{Schr\"odinger~}
\def\be#1\ee{\begin{equation}#1\end{equation}}
\def\ba#1\ea{\begin{align}#1\end{align}}
\def\bg#1\eg{\begin{gather}#1\end{gather}}
\def\t{\text}

\newcommand{\ketbra}[2]{
    \ensuremath{\ket{#1}\hspace{-2.5pt}\bra{#2}}
}
\newcommand{\selfketbra}[1]{
    \ketbra{#1}{#1}
}

\newcommand{\thermalstate}{
    \ensuremath{\rho_{\mathrm{th}}}
}

\newcommand{\heff}{
    \ensuremath{H_\mathrm{eff}}
}
\newcommand{\su}[1]{
    \ensuremath{\mathrm{SU}(#1)}
}

\newcommand{\givensRotation}[3]{
    \ensuremath{
        R\left(#1\right)_{#2}^{#3}
    }
}

\makeatletter
\newcounter{siequation}
\newcounter{sitable}
\newcounter{sifigure}
\@addtoreset{equation}{siequation}
\@addtoreset{table}{sitable}
\@addtoreset{figure}{sifigure}
\makeatother

\begin{document}

\title{Characterization of Control in a Superconducting Qutrit Using Randomized Benchmarking}

\author{M. Kononenko}
\email{mkononen@uwaterloo.ca}
\affiliation{
    Institute for Quantum Computing, University of
    Waterloo, Waterloo, ON, Canada, N2L 3G1
}
\affiliation{
    Department of Physics and Astronomy, University of
    Waterloo, Waterloo, ON, Canada, N2L 3G1
}

\author{M.A. Yurtalan}
\email{mayurtalan@uwaterloo.ca}
\affiliation{
    Institute for Quantum Computing, University of
    Waterloo, Waterloo, ON, Canada, N2L 3G1
}
\affiliation{
    Department of Physics and Astronomy, University of
    Waterloo, Waterloo, ON, Canada, N2L 3G1
}

\author{S. Ren}
\affiliation{
    Institute for Quantum Computing, University of
    Waterloo, Waterloo, ON, Canada, N2L 3G1
}
\affiliation{
    Department of Physics and Astronomy, University of
    Waterloo, Waterloo, ON, Canada, N2L 3G1
}

\author{J. Shi}
\affiliation{
    Institute for Quantum Computing, University of
    Waterloo, Waterloo, ON, Canada, N2L 3G1
}
\affiliation{
    Department of Physics and Astronomy, University of
    Waterloo, Waterloo, ON, Canada, N2L 3G1
}

\author{S. Ashhab}
\affiliation{Qatar Environment and Energy Research Institute, Hamad Bin Khalifa University, Qatar Foundation, Qatar}
\affiliation{Advanced ICT Research Institute, National Institute of Information and Communications Technology (NICT), 4-2-1, Nukui-Kitamachi, Koganei, Tokyo 184-8795, Japan}

\author{A. Lupascu}
\email{alupascu@uwaterloo.ca}
\affiliation{
    Institute for Quantum Computing, University of
    Waterloo, Waterloo, ON, Canada, N2L 3G1
}
\affiliation{
    Department of Physics and Astronomy, University of
    Waterloo, Waterloo, ON, Canada, N2L 3G1
}
\affiliation{Waterloo Institute for Nanotechnology, University
of Waterloo, Waterloo, ON, Canada N2L 3G1}

\date{ \today}

\begin{abstract}
We characterize control of a qutrit implemented in the lowest three energy levels of a capacitively-shunted flux-biased superconducting circuit. Randomized benchmarking over the qutrit Clifford group yields an average fidelity of $98.89 \pm 0.05 \, \%$. For a selected subset of the Clifford group, we perform quantum process tomography and observe the behaviour of repeated gate sequences. Each qutrit gate is generated using only two-state rotations via a method applicable to any unitary.  We find that errors are due to decoherence primarily and have a significant contribution from level shifts. This work demonstrates high-fidelity qutrit control and outlines avenues for future work on optimal control of superconducting qudits.
\end{abstract}

\maketitle

Recent advances in large-scale quantum information processors have relied on manipulating quantum information using two-level systems as qubits~\cite{aruteQuantumSupremacyUsing2019, otterbachUnsupervisedMachineLearning2017a, corcolesChallengesOpportunitiesNearTerm2019, wrightBenchmarking11qubitQuantum2019}. Theoretical work shows that using multilevel systems as qudits offers performance advantages in quantum error correction~\cite{campbellEnhancedFaultTolerantQuantum2014, anwarFastDecodersQudit2014, krishnaLowOverheadMagic2019, prakashMagicStateDistillation2020}, quantum sensing~\cite{shlyakhovQuantumMetrologyTransmon2018, suslovQuantumAbacusCounting2011}, and quantum communication~\cite{bouchardHighdimensionalQuantumCloning2017}. Efficient universal qudit control for these applications follows from an extension of the Solovay-Kitaev theorem from the qubit unitary group $\su{2}$ to the qudit group $\su{d}$, where $d$ is the dimension of the qudit's Hilbert space~\cite{dawsonSolovayKitaevAlgorithm2006}. However, implementation of such control brings new challenges including mapping qudit gates to experimentally accessible controls and understanding how control errors translate into errors in the qudit gates. Characterizing qudit gates is also more resource-intensive than characterizing qubit gates because the larger Hilbert space allows more complex states to form.

In this Letter, we characterize qutrit control using randomized benchmarking (RB), which is a protocol that yields the average fidelity for the elements of the Clifford group. We implement qutrit gates using a universal decomposition method that can be used to generate any unitary for a qutrit and more generally for qudits of any dimension. The measured average fidelity is $\mathcal{\bar{F}} = 98.89 \pm 0.05 \%$ for members of the qutrit Clifford group $\mathcal{C}_3$. In addition, we characterize a subset of the Clifford group using quantum process tomography (QPT), which provides an independent verification of the unitary synthesis method and yields results in agreement with RB. Analysis of this experiment revealed errors due to decoherence, level shifts, and leakage. Level shifts, which do not lead to significant errors in the usual case of resonant control in a two-dimensional subspace, are a significant source of error in this experiment, pointing to relevant future work on shaped control pulses for qudits.

\begin{figure}
    \includegraphics[width=\linewidth]{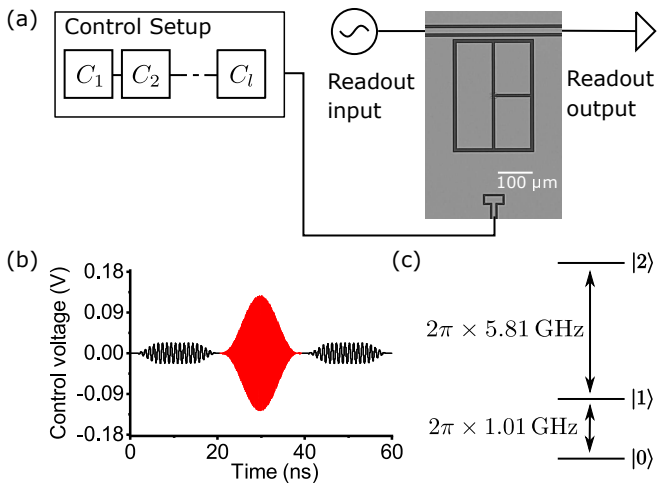}
    \caption{
        (a) A representation of the experiment setup showing a scanning electron micrograph of the device, capacitively coupled to a control line and to a readout resonator. The control setup synthesizes a sequence of Clifford gates $C_1$ to $C_l$ for an RB experiment. (b) The control waveform for a qutrit Walsh-Hadamard gate $H_3$. Black (red) lines are signals resonant with the $0-1$ ($1-2$) transition. Control voltage refers to the voltage at the output of the control setup. (c) The level diagram for the qutrit, with transition frequencies shown.
    }
    \label{fig:gate_synthesis}
\end{figure}

The experiment is performed on a variant of the capacitively-shunted flux qubit that combines relatively long coherence times with high anharmonicity~\cite{yurtalanCharacterizationMultilevelDynamics2020}. Anharmonicity here is defined as the difference between the second and first transition frequencies. A qutrit is encoded in the lowest three energy states of the device, denoted by $0$, $1$, and $2$. The large anharmonicity enables fast selective driving of the $0-1$ and $1-2$ transitions. Qutrit control is done using microwave pulses sent via a coplanar waveguide capacitively coupled to the device (see Fig. \ref{fig:gate_synthesis}(a)). Application of a microwave pulse resonant with the $m-n$ transition implements a rotation $R(\theta)_\phi^{n m}$ in the two dimensional space formed by states $m$ and $n$, where $\theta$ is the rotation angle and $\phi$ is the rotation axis phase (see Supplemental Material). Control pulses are synthesized using mixers Marki IQ1545 and IQ0307, for the $0-1$ and $1-2$ transitions respectively, fed by two Agilent E8527D PSG signal generators and quadrature control signals from a Tektronix AWG5014 arbitrary waveform generator. The device state is measured using heterodyne readout of a coplanar waveguide resonator also coupled capacitively to the device. The readout voltage, averaged over many repetitions, corresponds to the expectation value of the operator $V = V_0 \selfketbra{0} + V_1 \selfketbra{1} + V_2 \selfketbra{2}$. State preparation is done by waiting for the device to relax to the thermal state $\rho_\text{th} = P_{\t{th}, 0} \selfketbra{0} + P_{\t{th}, 1} \selfketbra{1}$, where $P_{\t{th}, n} = \braket{n | \thermalstate | n}$. We assume that higher state populations are negligible, in line with the large transition frequency between states 1 and 2. The thermal excited state population $P_{\t{th}, 1}$ and the readout voltage levels $V_0$, $V_1$, and $V_2$ are measured using a protocol based on observing Rabi oscillations on the $0-1$ transition starting with initial states based on the thermal state with additional population swaps (see Supplemental Material and Ref.~\cite{yurtalanCharacterizationMultilevelDynamics2020}). Based on the measured values of $V_0$, $V_1$, and $V_2$, the populations in any state can be characterized by measuring $\braket{V}$ in combination with suitable pre-measurement analyzer pulses (see Supplemental Material).

Performing a qudit gate $U$ requires a decomposition of $U$ that can be mapped to the available controls, which becomes more difficult as $d$ increases. In this experiment, each gate $U$ is decomposed into the product of a set of Givens rotations and a diagonal unitary, using an approach that is universal for any qudit of dimension $d$ as long as $d-1$ different transitions involving all the $d$ states can be controlled. The Givens rotations are determined based on the procedure in Ref.~\cite{schirmerConstructiveControlQuantum2002}; a maximum number of $\frac{1}{2} d (d - 1)$ Givens rotations are required. Each Givens rotation $R(\theta)_\phi^{mn}$ is then mapped to a control pulse with envelope area proportional to $\theta$, frequency resonant with the $m-n$ transition, and phase given by $\phi$. The control Hamiltonian for $R(\theta)_\phi^{mn}$ in the rotating frame is $H_\mathrm{drive}(t) = \frac{1}{2} \Omega_{mn}(t) e^{-i \phi} \ketbra{m}{n} + \mathrm{h.c.}$, where $\Omega_{mn}(t)$ is the drive strength for the $m-n$ transition. We use Givens rotations in subspaces $0-1$ and $1-2$. Rotations in the $0-1$ subspace have cosine-shaped rise and fall envelopes with a total duration of $18.4 \, \mathrm{ns}$. Rotations in the $1-2$ subspace have derivative removal by adiabatic gate (DRAG)~\cite{motzoiSimplePulsesElimination2009} cosine-shaped pulses with a duration of $16.8 \, \mathrm{ns}$ (see Supplemental Material). We use DRAG for the $1-2$ transition in order to suppress leakage to state $3$ arising from the small difference between the $1-2$ and $2-3$ transition frequencies ~\cite{chenMeasuringSuppressingQuantum2016}. The diagonal component of $U$ is tracked, and the phases of the Givens rotations in the subsequent gate are modified in order to implement phase gates at the software level \cite{yurtalanCharacterizationMultilevelDynamics2020, mckayEfficientGatesQuantum2017} (see Supplemental Material).

We verify this approach of implementing qutrit gates using QPT, which is a standard technique for finding the process matrix of a black box~\cite{chuangPrescriptionExperimentalDetermination1997}. We implement QPT using the procedure from Ref.~\cite{yurtalanImplementationWalshHadamardGate2020}. The gate fidelity~\cite{nielsenSimpleFormulaAverage2002a} calculated using the process matrix determines whether a qutrit gate is synthesized as intended. We performed QPT for two representative gates - the Walsh-Hadamard gate $H_3$, which together with the phase gate $S_3$ generates the Clifford group, and the Pauli $X_3$ gate, which together with the Pauli $Z_3$ gate generates the Pauli group~\cite{glaudellCanonicalFormsSinglequtrit2019, jafarzadehRandomizedBenchmarkingQudit2020}. We note that $S_3$ and $Z_3$ are diagonal gates, and are therefore software defined. The process fidelity for the Walsh-Hadamard gate $H_3$ is $97.45\%$ and the generalized Pauli gate $X_3$ is $98.47\%$. QPT shows that the experimental implementation of the gate decomposition synthesizes the intended gates.

\begin{figure}
    \centering
    \includegraphics[width=\linewidth]{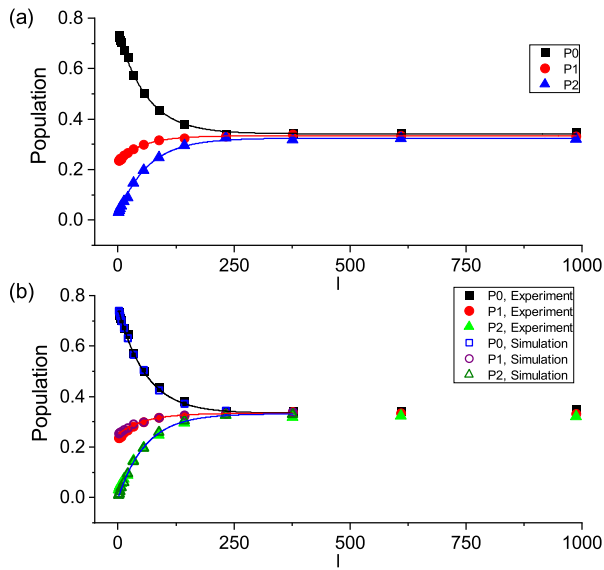}
    \caption{(a) The experimentally-measured average populations $P_0$ (black squares), $P_1$ (red circles), and $P_2$ (blue triangles) versus sequence length $l$. Solid lines show exponential fits. For $P_0$, $P_1$, and $P_2$, the $p_n$ values are $p_0 = 0.9839 \pm 5.4 \times 10^{-4}$, $p_1 = 0.9814 \pm 1.1 \times 10^{-3}$, and $p_2 = 0.9846 \pm 6.9 \times 10^{-4}$, respectively. (b) The experimentally-measured $P_0$ (black squares), $P_1$ (red circles), and $P_2$ (green triangles) versus the sequence length $l$, compared to the simulated $P_0$ (hollow blue squares), $P_1$ (hollow purple circles) and $P_2$ (hollow dark green triangles). Solid lines show the fits of the RB model to the simulated $P_0$, $P_1$, and $P_2$.}
    \label{fig:rb_results}
\end{figure}

Next, the average fidelity $\bar{\mathcal{F}}$ of $\mathcal{C}_3$ is characterized using RB~\cite{jafarzadehRandomizedBenchmarkingQudit2020}. RB relies on the fact that the application of a sequence of $l$ gates, of which the first $l - 1$ are chosen randomly from $\mathcal{C}_3$ and the last is chosen as the inverse of the product of the first $l - 1$ gates, behaves on average as a depolarizing channel with depolarizing coefficient $p^l$~\cite{emersonScalableNoiseEstimation2005b, magesanScalableRobustRandomized2011, jafarzadehRandomizedBenchmarkingQudit2020}. Specifically, the population $P_n$ for each level $n$ decays with $l$ as
\be
\label{eqn:rb_model}
P_n = (P_{\text{i} n} - 1/3) p^l + 1/3,
\ee
where $P_{\text{i} n}$ is the initial population of state $n$. The average gate fidelity $\bar{\mathcal{F}}$ is related to the depolarization coefficient $p$ via $\bar{\mathcal{F}} = p + (1 - p)/3$.
Randomized benchmarking is more efficient than QPT for measuring $\bar{\mathcal{F}}$ while being immune to state preparation and measurement errors, at the expense of not giving information about individual gates in $\mathcal{C}_3$~\cite{magesanScalableRobustRandomized2011}. Based on the fact that $H_d$ and $S_d$ generate $\mathcal{C}_d$ for any prime $d$~\cite{gottesmanFaultTolerantQuantumComputation1999}, we find all $216$ elements of $\mathcal{C}_3$ by calculating all distinct products of $H_3$ and $S_3$. Each Clifford gate is decomposed into Givens rotations as explained above. We measure the result of applying a set of random gates of length $l$ ranging from $l = 2$ to $l = 987$, with $25$ different randomizations for each value of $l$. The measurement is repeated $N_\text{rep} = 8192$ times for each different random sequence. Figure \ref{fig:rb_results}(a) shows the results of a RB experiment, with the initial state set to be the thermal state.  The fit of the populations versus the sequence length, given by $P_n = (P_{\text{i} n} - P_{\text{f} n}) p_n^l + P_{\text{f} n}$, with $P_{\text{i} n}$, $P_{\text{f} n}$ and $p_n$ as free parameters, is in excellent agreement with the data. The polarization decay coefficients are the same within the experimental errors, with an average $p = 0.9833 \pm 5 \times 10^{-4}$ and a corresponding fidelity $\bar{\mathcal{F}} = 98.89 \pm 0.05 \, \%$. Similarly the final values $P_{\text{f} 0} = 0.341 \pm 0.003$, $P_{\text{f} 1} = 0.333 \pm 0.001$, and $P_{\text{f} 2} = 0.325 \pm 0.003$ are close to the expected value of $\frac{1}{3}$ corresponding to the fully depolarized state of a qutrit. Repeating the experiment with $50$ randomizations for each $l$ gave the same result to within experimental error.

\begin{figure}
    \centering
    \includegraphics[width=\linewidth]{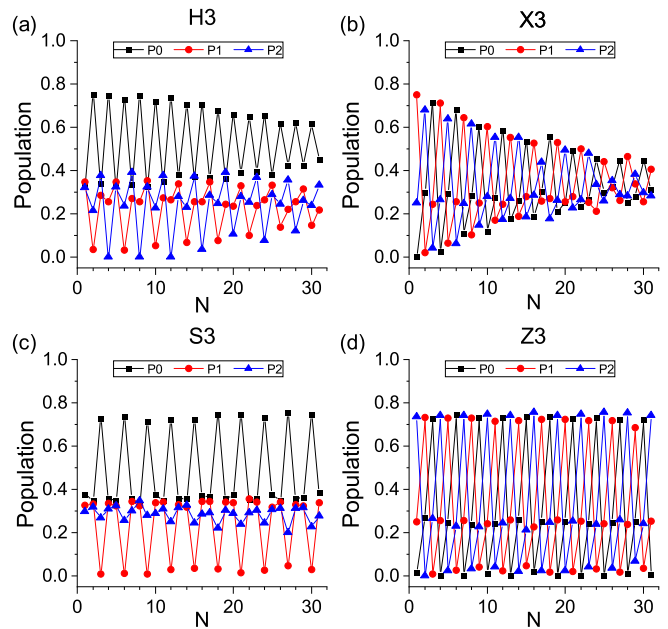}
    \caption{The experimentally-measured populations $P_0$ (black squares), $P_1$ (red circles), and $P_2$ (blue triangles) versus the number $N$ of repetitions of gates $H_3$ (a), $X_3$ (b), $S_3$ (c), and $Z_3$ (d).}
    \label{fig:zigzag_results}
\end{figure}

Since each element in $\mathcal{C}_3$ has a finite order, repeated application of the same gate leads to a periodic result versus the number of repetitions $N$. This is tested experimentally by measuring the populations after repeated application of $H_3$, $S_3$, $X_3$, and $Z_3$, starting with the thermal state as an initial state. For $S_3$ and $Z_3$, an $H_3$ gate is prepended and appended to produce a change in the measured populations. Figure \ref{fig:zigzag_results} shows the populations $P_n$ for these gates. The experiments confirm the expected periodicity of these elements of the Clifford group, which is $4$, $3$, $3$, and $3$ for $H_3$, $S_3$, $X_3$, and $Z_3$, respectively. The effect of control errors and decoherence is visible as $N$ increases, in good agreement with numerical simulations of the dynamics (see Supplemental Material). The lack of decay in $S_3$ and $Z_3$ signals is due to these gates being entirely defined in software following the decomposition above. In the $S_3$ and $Z_3$ sequence, errors in control arise only from the preparation and post $H_3$ pulses.

We now discuss the sources of error in the RB experiment. Numerical simulations of the dynamics are done with two models, based on coherent and incoherent evolution. The device Hamiltonian is truncated to the seven lowest energy states, which was found to be sufficient to properly explain level shifts in previous work including two-photon driving (See Supplemental Material and Ref.~\cite{yurtalanImplementationWalshHadamardGate2020}). The superconducting circuit parameters are extracted based on a model fit against the measured spectroscopy data. In addition, calibration of the pulses is done in the simulation in a manner similar to the experiment (see Supplemental Material). We model the dynamics under the randomized benchmarking sequences used in the experiment, by numerically solving the \sch equation with driving~\cite{johanssonQuTiPPythonFramework2013}. With coherent evolution, state populations extracted from the simulations give a fidelity of $\bar{\mathcal{F}} = 99.91 \pm 0.02 \%$, significantly higher that the experimental result. In addition, simulations indicate that leakage to states outside the qutrit space is negligibly small ($2.29 \times 10^{-4}\,\%$ at $l = 377$).

The role of decoherence is analyzed using a model that includes the measured relaxation and excitation rates and dephasing in the qutrit space.  The decoherence model consists of Lindblad operators of the form $\sqrt{\Gamma_1^{mn}} \ketbra{n}{m}$ for relaxation/excitation, and $\sqrt{\frac{1}{2} \Gamma_2^{mn}} (\ketbra{m}{m} - \ketbra{n}{n})$ for dephasing (see Supplemental Material). Dynamics are simulated using a master equation solver~\cite{nielsenQuantumComputationQuantum2013, johanssonQuTiPPythonFramework2013}. Figure \ref{fig:rb_results}(b) shows the numerically calculated populations with decoherence. The fit of this simulation gives a fidelity of $\bar{\mathcal{F}} = 98.9 \pm 0.03 \%$, which is within the fit error of the experimental result. As for the coherent case, the total population outside the qutrit space is negligible, reaching $0.244\%$ at $l = 377$. Numerical simulation of each gate in $\mathcal{C}_3$ individually, and then averaging their respective fidelities, gives $\bar{\mathcal{F}} = 98.9 \, \%$ with a standard deviation of $0.3 \, \%$ and a worst-case fidelity of $98.5 \, \%$. This compares well with experimental values, and indicates that in addition to the average fidelity being high, each gate in $\mathcal{C}_3$ is synthesized with high fidelity as well. The range in fidelity is also comparable to the fidelities measured using QPT. From fitting the model in Ref. \cite{woodQuantificationCharacterizationLeakage2018} to the numerically calculated populations, we determine the leakage $L_1$ per Clifford gate \cite{chenMeasuringSuppressingQuantum2016}, the seepage $L_2$, and the adjusted average fidelity $\bar{\mathcal{F}}_L$. Whereas leakage measures population transfer out of the qutrit space, seepage measures population transfer into the qutrit space. The simulation yields $L_1 = 6.49 \times 10^{-7} \pm 6.2 \times 10^{-5}$, $L_2 = 7.54 \times 10^{-7} \pm 6.70 \times 10^{-5}$, and $\bar{\mathcal{F}}_L = 98.9 \pm 0.2 \%$. Since $\bar{\mathcal{F}}_L$ is not significantly smaller than the simulated $\bar{\mathcal{F}}$, leakage does not contribute significantly to the error in the simulation.

The analysis discussed above indicates that the dominant source of error in the current experiment is decoherence. Faster control using stronger driving can help to mitigate this error, but coherent control errors must be understood, since these errors increase with faster driving. To analyze coherent control errors in the qutrit RB experiment, it is useful to connect the errors of Clifford group unitaries to errors in their component $R(\theta)_\phi^{mn}$. For a noisy implementation $\tilde{C} = \prod\limits_n \tilde{R}_n$ of a qutrit gate $C = \prod\limits_n R_n$, with $\tilde{R}_n$ of noisy versions of ideal Givens rotations $R_n$, the gate error $r(\tilde{C}, C) = 1 - \mathcal{F}(\tilde{C}, C)$ is approximately $r(\tilde{C}, C) \approx \sum\limits_n r(\tilde{R}_n, R_n)$, with $r(\tilde{R}_n, R_n)$ the error for a Givens rotation. The approximation follows from modeling the error for a Givens rotation as an operator $K_n = \tilde{R_n} R_n^\dagger$, with $K_n = \alpha_n I + \beta_n M_n$, where $\alpha_n$ the complex number minimizing $||I_3(K_n - \alpha_n I)I_3||_\infty$ and $I_3 = \selfketbra{0} + \selfketbra{1} + \selfketbra{2}$, $\beta_n = ||I_3(K - \alpha I_n)I_3||_\infty$, and $M_n = \frac{1}{\beta_n} (K_n - \alpha_n I)$. $||A||_\infty$ is the magnitude of the largest eigenvalue for an operator $A$. Relating $\alpha_n$ and $\beta_n$ to $\mathcal{F}(\tilde{R}_n, R_n)$ and $\mathcal{F}(\tilde{C}, C)$ using
\be
\mathcal{F}(\tilde{A}, A) = \frac{
    \sum\limits_{U_j \in \mathcal{P}_d} \trace{}{
        A^\dagger U_j A \tilde{A}U_j\tilde{A}^\dagger}
    + d^2}{d^2 (d + 1)}
\label{eqn:nielsen_formula}
\ee
for two unitary operators $A$ and $\tilde{A}$ yields $r(\tilde{C}, C) \approx \sum\limits_n r(\tilde{R}_n, R_n)$, assuming products of $\beta_n$ are small (see Supplemental Material). Figure \ref{fig:givens_results}(a) shows that the approximation holds numerically. 

\begin{figure}[!ht]
    \centering
\includegraphics[width=\linewidth]{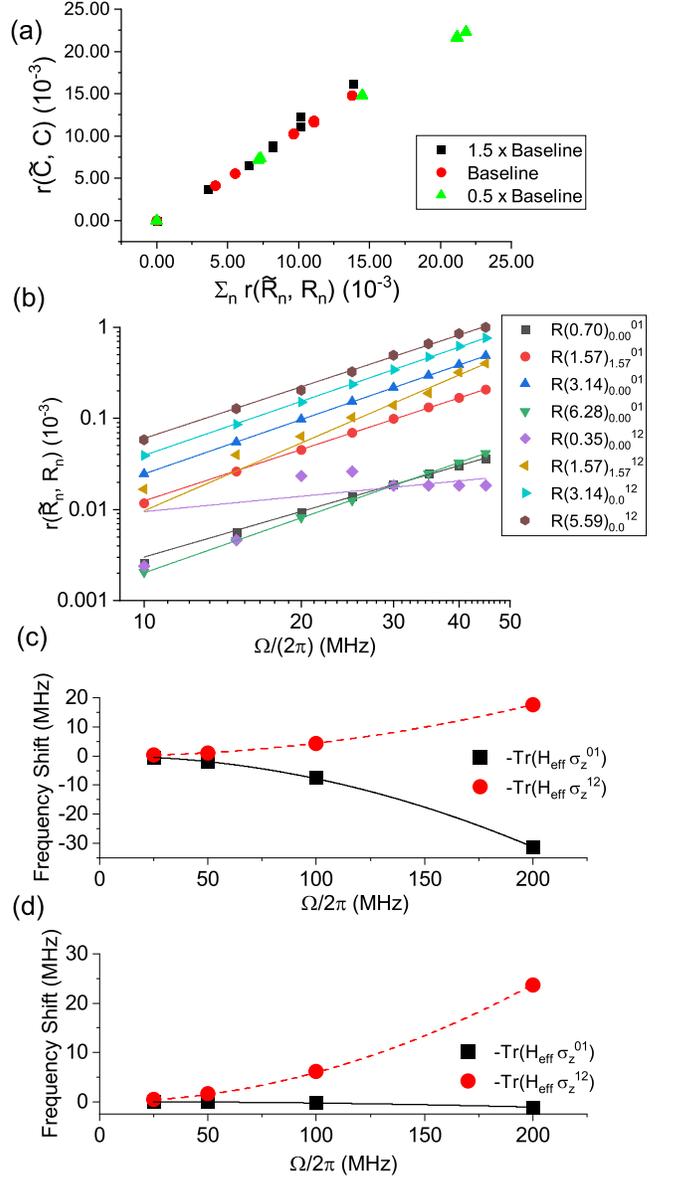}
    \caption{
        (a) The error of the Clifford gate versus the sum of the errors of its constituent Givens rotations for all gates in $\mathcal{C}_3$. The black squares, red circles, and green triangles correspond to amplitudes multiplied by 1.5, 1.0 and 0.5 relative to the experiment pulses respectively. Pulse durations for each multiplication case are adjusted to preserve the intended rotation angles. (b) The error $r = 1 - \mathcal{F}$ versus the drive strength for a set of Givens rotations. Different symbols correspond to the different rotations. (c, d) The value of the indicated component of the effective Hamiltonian for $R(\pi)_0^{01}$ (c) and $R(\pi)_0^{12}$ (d) versus drive strength. Black squares (red circles) indicate the $01$ ($12$) level shift. The black solid (red dashed) line shows a quadratic fit to the $01$ ($12$) level shift versus drive frequency.
    }
    \label{fig:givens_results}
\end{figure}

To understand the errors in individual Givens rotations, the effective Hamiltonian $\heff = -\frac{i}{\tau} \ln \tilde{R}_n$ is calculated for each numerically calculated $\tilde{R}_n$, where $\tau$ is the effective time to implement $R_n$ and $\ln \tilde{R}_n$ is the matrix logarithm. The difference between $\heff$ and the ideal control Hamiltonian $H_\t{drive}$ can be connected to errors introduced by the failure of the rotating wave approximation. Figure \ref{fig:givens_results}(b) shows the error of the Givens rotations versus drive strength, showing that the error scales quadratically with the drive strength. The effective Hamiltonian is well-approximated by $H_\text{drive} + \sum\limits_{m, n} s_{mn} \Omega_{mn}^2 \sigma_z^{mn}$, with $\sigma_z^{mn} = \selfketbra{m} - \selfketbra{n}$, and $s_{mn}$ being a real coefficient. The quadratic scaling indicates that the main errors are caused by driving-induced level shifts. Figure \ref{fig:givens_results}(c) and (d) show the total  frequency shifts versus the drive strength, i.e.  $\trace{}{-\heff \sigma_z^{mn}}$, for $mn = 01$ and $mn = 12$ transitions induced by $R(\pi)_0^{01}$ and $R(\pi)_0^{12}$ pulses, respectively. The calculated total frequency shifts are in excellent agreement with the analytical formula given in Ref.~\cite{yurtalanImplementationWalshHadamardGate2020}. The leakage error identified in the RB analysis is also visible in simulation of the Givens rotations as non-negligible values of $\braket{2 | \heff | 3}$, $\braket{1 | \heff | 6}$, and $\braket{2 | \heff | 6}$. However, these terms are small compared to $\trace{}{\heff \sigma_z^{mn}}$, confirming that contribution of leakage to the control error is small compared to level shifts. Note that level shift errors are much more significant for qutrit control than for qubit control, since the level shift is a coherent error on levels used to store information in qutrits, compared to an incoherent error on levels not used to store information in qubits.

In conclusion, we demonstrated control sufficient to synthesize the qutrit Clifford group $\mathcal{C}_3$ with $98.89 \pm 0.05 \%$ fidelity, using a universal method for gate decomposition into Givens rotations. This fidelity is in agreement with QPT data. The experimental errors are dominated by decoherence, but level shifts due to off-resonant coupling to states outside the driven two-dimensional subspace of each Givens rotation contribute to control error as well. In future work it will be interesting to explore the application of level shift corrections, as done in Ref.~\cite{yurtalanImplementationWalshHadamardGate2020}, and more generally design optimal control pulses that mitigate both level shifts and leakage. These results establish randomized benchmarking as a tool to understand superconducting qutrit control and pave the way towards using superconducting qudits in quantum information tasks.

We thank the University of Waterloo Quantum Nanofab team members for their help during device fabrication. We acknowledge support from NSERC, the Canada Foundation for Innovation (CFI), the Ontario Ministry of Research and Innovation, Industry Canada, and the Canadian Microelectronics Corporation (CMC). S.A. was supported in part by the Ministry of Education, Culture, Sports, Science and Technology (MEXT) Quantum Leap Flagship Program Grant Number JPMXS0120319794.
M.K and A.Y. contributed equally to the work.

\textit{Note added:} During the completion of our manuscript, we became aware of a recent related manuscript in qutrit randomized benchmarking ~\cite{morvanQutritRandomizedBenchmarking2020}.

\onecolumngrid
\clearpage
\begin{center}
	\textbf{\large {Supplemental Material for ``Characterization of Control in a Superconducting Qutrit Using Randomized Benchmarking''}}
\end{center}

\setcounter{page}{1}
\stepcounter{siequation}
\stepcounter{sifigure}
\stepcounter{sitable}
\makeatletter
\renewcommand{\theequation}{S\arabic{equation}}
\renewcommand{\thefigure}{S\arabic{figure}}
\renewcommand{\thetable}{S\arabic{table}}
\renewcommand{\bibnumfmt}[1]{[S#1]}
\renewcommand{\citenumfont}[1]{S#1}

\section{Experiment Details}
\label{si:sec:experiment_details}

The device is manufactured using a planar process described in Ref.~\cite{SIyurtalanImplementationWalshHadamardGate2020} (Supplemental Material). The qutrit is capacitively coupled to a coplanar waveguide resonator for dispersive readout, and to a transmission line terminated by a capacitor for control. The device is placed in a microwave package and mounted in a dilution cryostat. Experiments are done with the device operated at its flux symmetry point, with the required flux provided by an external coil. The measured transition frequencies are $\omega_{01} = 2 \pi \times 1.01 \, \mathrm{GHz}$ and $\omega_{12} = 2 \pi \times 5.81 \, \mathrm{GHz}$, where $\omega_{mn}$ is the frequency of the transition between states $m$ and $n$.  Readout is done with a resonator with frequency $2 \pi \times 6.72 \, \mathrm{GHz}$ and a quality factor $Q = 1.06 \times 10^5$, coupling to the lowest two levels of the device with a Jaynes-Cummings interaction strength $g = 60 \, \mathrm{MHz}$.

The populations of the thermal state $\thermalstate$ are measured by comparing the amplitude of two Rabi oscillations induced on the $0-1$ transition. The first Rabi oscillation uses the thermal state, followed by a $1-2$ population swap, as the initial state. The second Rabi oscillation uses the thermal state followed by $0-1$ and $1-2$ population swaps as the initial state. Here, it is assumed that in the thermal state the populations of state 2 and higher energy states are negligible, in line with the large 1-2 transition frequency. The voltage levels $V_0$, $V_1$, and $V_2$ in the readout voltage observable $V$ are characterized by measuring $\braket{V}$ for three reference states: the thermal state, the thermal state followed by swapping states $0$ and $1$, and the thermal state followed by swapping states $1$ and $2$, then swapping states $0$ and $1$~\cite{SIyurtalanCharacterizationMultilevelDynamics2020}.  From these measurements, $\rho_{\mathrm{th}} = (0.753 \pm 0.003) \selfketbra{0} + (0.247 \pm 0.003) \selfketbra{1}$, $V_0 = 208 \pm 6 \, \mathrm{\mu V}$, $V_1 = -653 \pm 4 \, \mathrm{\mu V}$, and $V_2 = -392 \pm 6 \, \mathrm{\mu V}$. The ground state population of the thermal state ${\langle 0|\thermalstate|0\rangle}$ is consistent with an effective temperature of $43.6 \pm 0.7 \, \mathrm{mK}$.

Control pulses are expressed in the form $R(\theta)_\phi^{mn}$, where $\theta$ is the angle of rotation, $\phi$ is the phase, and $mn$ indicates the state $m$ to state $n$ transition being driven. To experimentally measure populations, i.e. the diagonal components of a given density matrix $\rho$, three measurements of $\braket{V}$, labelled $M_1$, $M_2$, and $M_3$ were conducted with various population swap pulses applied just before the measurement. $M_1$ is measured without any population swap pulses. $M_2$ is measured with a $R(\pi)_0^{01}$ pulse, and $M_3$ is measured  with $R(\pi)_0^{01} R(\pi)_0^{12} R(\pi)_0^{01}$ pulses applied.  The measured quantities $M_1$, $M_2$ and $M_3$ correspond to
\begin{equation}
M_1 = \trace{}{V \rho},
\end{equation}
\begin{equation}
M_2 = \trace{}{V R(\pi)_0^{01} \rho {R(\pi)_0^{01}}^\dagger},
\end{equation}
and
\begin{equation}
M_3 = \trace{}{V R(\pi)_0^{01} R(\pi)_0^{12} R(\pi)_0^{01} \rho {(R(\pi)_0^{01} R(\pi)_0^{12} R(\pi)_0^{01})}^\dagger}.
\end{equation}
The populations are then obtained by finding $p_0$, $p_1$ and $p_2$ satisfying $0 \leq p_0 \leq 1$, $0 \leq p_1 \leq 1$, $p_0 + p_1 + p_2 = 1$, and minimizing the objective function
\begin{equation}
    f(p_0, p_1, p_2) = \left|\left|
        \left(\begin{array}{c} M_1 \\ M_2 \\ M_3\end{array}\right) - \left(\begin{array}{ccc}
            V_0 & V_1 & V_2 \\ V_1 & V_0 & V_2 \\ V_2 & V_1 & V_0
        \end{array}\right) \left(\begin{array}{c}p_0\\p_1\\p_2\end{array}\right)
    \right|\right|,
\end{equation}
where $||x||$ is the Euclidean norm of $x$.

The pulse calibration procedure for $\frac{\pi}{2}$ rotations around the $x$ and $y$ axes, denoted $R\left(\frac{\pi}{2}\right)_0^{01}$ and $R\left(\frac{\pi}{2}\right)_{\frac{\pi}{2}}^{01}$ respectively, was done by applying the rotation $2m+1$ times to the qutrit thermal state , where $m$ is an integer. The measured readout voltage $\braket{V}$ versus $2m+1$ was then used to determine the  error in the rotation angle and correct for the pulse amplitude iteratively. Next, the phase of $\givensRotation{\frac{\pi}{2}}{\frac{\pi}{2}}{01}$ relative to $\givensRotation{\frac{\pi}{2}}{0}{01}$ was calibrated by measuring $\braket{V}$ versus $m$ for the sequence
\begin{equation}
\givensRotation{\frac{\pi}{2}}{0}{01} \left(
    \left(\givensRotation{\frac{\pi}{2}}{\frac{\pi}{2}}{01}\right)^2
    \left(\givensRotation{\frac{\pi}{2}}{0}{01}\right)^2
\right)^m
\givensRotation{\frac{\pi}{2}}{\frac{\pi}{2}}{01}.
\end{equation}
 After these corrections, arbitrary rotations $R(\theta)_\phi^{01}$ are then implemented by interpolating between the calibrated rotations. To implement an arbitrary $\theta$, the pulse amplitude is scaled by a factor of $\frac{2 \theta}{\pi}$ relative to the calibrated pulses, and to implement an arbitrary $\phi$, the pulse phase is adjusted using the phase of the calibrated $\givensRotation{\frac{\pi}{2}}{0}{01}$ and $\givensRotation{\frac{\pi}{2}}{\frac{\pi}{2}}{01}$ as a reference. A similar procedure is used to calibrate the rotations $R(\theta)_\phi^{12}$.

We extract the coherence times of the device used as a qutrit following Ref.~\cite{SIyurtalanCharacterizationMultilevelDynamics2020}. Table \ref{si:tbl:gamma_factors} shows the relaxation ($m>n$) and excitation ($m<n$) rates $\Gamma_{1}^{mn}$ and the Ramsey dephasing rates $\Gamma_{2}^{mn}$ for all pairs of the qutrit energy levels~\cite{SIlevittSpinDynamicsBasics2008}.
\begin{table}
    \centering
    \caption{The decoherence rates measured in the qutrit space.}
    \label{si:tbl:gamma_factors}
    \begin{tabular*}{1.7in}{l @{\extracolsep{\fill} } r}  \hline \hline
         Rate & Value (Hz)  \\ \hline
         $\Gamma_1^{10}$ & $2.83 \times 10^4$ \\ 
         $\Gamma_1^{01}$ & $1.49 \times 10^3$ \\
         $\Gamma_1^{21}$ & $1.80 \times 10^5$ \\
         $\Gamma_1^{12}$ & $2.89 \times 10^2$ \\
         $\Gamma_1^{20}$ & $3.91 \times 10^4$ \\
         $\Gamma_1^{02}$ & $2.34 \times 10^1$ \\
         $\Gamma_2^{01}$ & $9.60 \times 10^4$ \\
         $\Gamma_2^{12}$ & $3.30 \times 10^5$ \\
         $\Gamma_2^{02}$ & $1.02 \times 10^5$ \\ \hline \hline
    \end{tabular*}
\end{table}

\section{Qutrit Gate Decompositions}
Following Ref.~\cite{SIschirmerConstructiveControlQuantum2002}, an arbitrary qutrit gate $U$ can be decomposed into the form
\begin{equation}
    \label{si:eqn:gate_decomp}
    U = U_d R(\theta_1)_{\phi_1}^{01} R(\theta_2)_{\phi_2}^{12} R(\theta_3)_{\phi_3}^{01},
\end{equation}
where $R(\theta)_\phi^{mn}$ are Givens rotations and the matrix $U_d$ is diagonal in the computational basis. Each $R(\theta)_\phi^{mn}$ is expressed by
\begin{equation}
    R(\theta)_\phi^{mn} = \exp\left(-i \frac{\theta}{2} \left(
        \cos(\phi) \sigma_x^{mn} + \sin(\phi) \sigma_y^{mn}
    \right)
    \right),
\end{equation}
where $\sigma_x^{mn} = \ketbra{m}{n} + \ketbra{n}{m}$ and $\sigma_y^{mn} = i (\ketbra{n}{m} - \ketbra{m}{n})$. The rotation $R(\theta)_\phi^{mn}$  can set the matrix element $\braket{m | U {R(\theta)_\phi^{mn}}^\dagger | a}$ to $0$, where $\ket{a}$ is a computational basis state, by setting
\begin{equation}
    \theta = 2 \arccos{\frac{|\braket{m | U | a}|^2}{
        \sqrt{|\braket{m | U | a}|^2 + |\braket{n | U | a}|^2}
        }
    }
\end{equation}
and
\begin{equation}
    \phi = \frac{\pi}{2} - \arg{\left(\braket{m | U | a}\right)} + \arg{\left(\braket{n | U | a}\right)}.
\end{equation}
Using this property, the rotations $R(\theta)_\phi^{mn}$ in Eq. (\ref{si:eqn:gate_decomp}) sequentially implement the off-diagonal elements of $U$ with  $\braket{0 | U \left(R(\theta_1)_{\phi_1}^{01}\right)^\dagger | 2}= 0$,   $\braket{1 | U \left(R(\theta_1)_{\phi_1}^{01} R(\theta_2)_{\phi_2}^{12}\right)^\dagger | 2}=0$ , and $\braket{0 | U \left( R(\theta_1)_{\phi_1}^{01} R(\theta_2)_{\phi_2}^{12} R(\theta_3)_{\phi_3}^{01} \right)^\dagger | 1} = 0$. If $\theta = 0$ for a particular rotation, then this rotation is omitted. After identifying the $\theta$ and $\phi$ for all the Givens rotations, the remaining diagonal part $U_d$ is implemented in software. This software implementation is done by shifting the phases of subsequent rotations  $R(\theta)_{\phi}^{mn}$  by $\arg{\left(\braket{m | U_d | m}\right)} - \arg{\left(\braket{n | U_d | n}\right)}$ \cite{SImckayEfficientGatesQuantum2017}.

\begin{figure}
    \centering
    \includegraphics{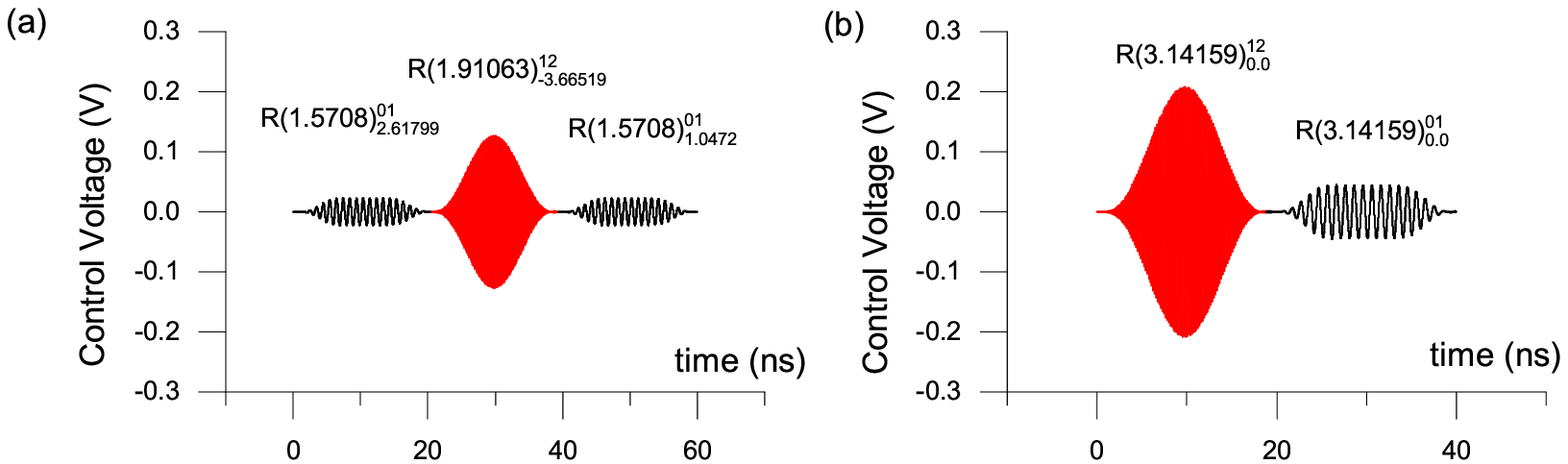}
    \caption{The waveforms implementing the qutrit gates (a) $H_3$, and (b) $X_3$. The control setup output voltage versus time is shown  for $01$ (Black lines) and $12$ (Red lines) transitions. The rotation implemented by each section of the waveform is as indicated.}
    \label{si:fig:gate_waveforms}
\end{figure}

Each $R(\theta)_\phi^{01}$ in Eq.(\ref{si:eqn:gate_decomp}) is mapped to a control pulse with a cosine rise and fall envelope with $t_\mathrm{rise} = t_\mathrm{fall} = 5 \, \mathrm{ns}$, and a total duration of $18.4 \, \mathrm{ns}$. Each $R(\theta)_\phi^{12}$ in Eq.(\ref{si:eqn:gate_decomp}) is mapped to a cosine derivative removal by adiabatic gate (DRAG) pulse~\cite{SImotzoiSimplePulsesElimination2009} with a duration of $16.8 \, \mathrm{ns}$. Figure \ref{si:fig:gate_waveforms} gives the control waveforms implementing the Walsh-Hadamard gate $H_3$  and the qutrit Pauli gate $X_3$. Following this decomposition, on average, $2.625$ Givens rotations per Clifford gate are required to implement the qutrit Clifford group $\mathcal{C}_3$. Generalizing the decomposition presented above from qutrit gates to a $d$-dimensional qudit gate $U$ requires a maximum of $\frac{1}{2}d (d - 1)$ rotations \cite{SIschirmerConstructiveControlQuantum2002}, which is the number of elements of $U$ in the upper-triangular region.

\section{Numerical Simulation of Device Dynamics}
\label{si:sec:numerical_simulations}

The circuit Hamiltonian, derived in the model presented in Ref.~\cite{SIyurtalanCharacterizationMultilevelDynamics2020}, together with the drive has the form $H(V_\text{q}(t)) = H_\mathrm{static} + H_\mathrm{drive}(V_\text{q}(t))$, where  $H_\mathrm{static} = \sum_{j=1}^{s-1} \omega_{0j} |j\rangle\langle j|$  and $ H_\mathrm{drive}(V_\text{q}(t)) = \sum_{\substack{i,j=0 \\ \substack{i\neq j}} }^{s-1} V_\text{q}(t)(g_{ij}|i\rangle\langle j| + \mathrm{h.c.})$ describes time-dependent control with $V_\text{q}(t)$ being the voltage at the control pad of the device. The Hilbert space is truncated to the lowest seven levels ($s=7$), owing to the agreement between the simulation and previous multi-level control experiments including two-photon Rabi oscillations~\cite{SIyurtalanImplementationWalshHadamardGate2020}. The transition frequencies $\omega_{0j}$  and the transition matrix elements $g_{ij}$ are calculated from the circuit model fitted to experimental data and are given in  Table \ref{si:tbl:omegas} and Table \ref{si:tbl:drive_coefficients} respectively.

\begin{table}
	\caption{The transition frequencies $\omega_{0j}$ in $H_\mathrm{static}$ used in simulations.}
	\label{si:tbl:omegas}
	\centering
	\begin{tabular*}{2.5in}{l @{\extracolsep{\fill}} r}
		\hline
		\hline
	    Coefficient & Value ($2 \pi \times 10^9 \, \mathrm{Hz}$) \\ \hline
	    $\omega_{01}$ & $1.005$ \\
	    $\omega_{02}$ & $6.837$ \\
	    $\omega_{03}$ & $11.81$ \\
	    $\omega_{04}$ & $17.17$ \\
	    $\omega_{05}$ & $17.63$ \\
	    $\omega_{06}$ & $18.16$ \\
		\hline
	\end{tabular*}
\end{table}

\begin{table}
	\caption{The transition matrix elements $g_{ij}$ of $H_\mathrm{drive}$ used in simulations. All other coefficients not shown are negligible.}
	\label{si:tbl:drive_coefficients}
	\centering
	\begin{tabular*}{2.5in}{l @{\extracolsep{\fill}} r}
		\hline
		\hline
	    Coefficient & Value ($2 \pi \times 10^9 \, \mathrm{Hz} \, \mathrm{V}^{-1}$) \\ \hline
	    $g_{01}$ & $99.1$ \\
	    $g_{03}$ & $-200.0 + 134.8i$ \\
	    $g_{04}$ & $67.11 + 17.13i$ \\
	    $g_{12}$ & $365.6$ \\
	    $g_{15}$ & $42.22 - 56.34i$ \\
	    $g_{16}$ & $144.5 - 79.97i$ \\
	    $g_{23}$ & $350.4 - 236.3i$ \\
	    $g_{24}$ & $-11.30 - 52.884i$ \\
	    $g_{35}$ & $-93.07 - 32.32i$ \\
	    $g_{36}$ & $-583.3 - 51.47i$ \\
	    $g_{45}$ & $20.41 - 49.20i$ \\
	    $g_{46}$ & $-2.069 + 1.948i$ \\
		\hline
	\end{tabular*}
\end{table}

The decoherence model consists of Lindblad operators of the form $\sqrt{\Gamma_1^{mn}} \ketbra{n}{m}$ for relaxation/excitation, and $\sqrt{\frac{1}{2} \Gamma_2^{mn}} (\ketbra{m}{m} - \ketbra{n}{n})$ for dephasing. To determine $\rho(t)$ after applying an RB sequence, the Lindblad master equation~\cite{SInielsenQuantumComputationQuantum2013} is solved numerically~\cite{SIjohanssonQuTiPPythonFramework2013} with $\thermalstate$ as the initial state. To determine the fidelity of individual Clifford gates and Givens rotations, each member $P$ of the three-level Pauli group $\mathcal{P}_3$ was propagated by numerically solving the Lindblad master equation with $P$ as the initial condition. The Pauli operator $P \in \mathcal{P}_3$ and the propagated Pauli operator $\Lambda(P)$, where $\Lambda$ the propagator for the device dynamics,  were then used to find the fidelity $\bar{\mathcal{F}}$. The control waveform voltage is related to the voltage at the device control pad $V_\text{q}(t)$ by the transfer coefficients $c_{01}$ and $c_{12}$. The coefficient $c_{01}$ is determined by simulating a $R\left(\frac{\pi}{2}\right)_0^{01}$ pulse, and setting $c_{01}$ such that the populations in $\ket{0}$ and $\ket{1}$ are equal after applying the simulated propagator to the initial state $\ketbra{0}{0}$. A similar procedure is done to calibrate $c_{12}$, but with $\ketbra{1}{1}$ as the initial state, and comparing the populations in $\ket{1}$ and $\ket{2}$.

\section{Calculation of Gate Fidelity}
\label{si:sec:gate_fidelity}

Following \cite{SInielsenSimpleFormulaAverage2002a}, the gate fidelity $\mathcal{F}(\mathcal{\tilde{U}}, U)$, where $\mathcal{\tilde{U}}$ is a noisy implementation of $U$, is given by
\begin{equation}
    \label{si:eqn:nielsen_formula}
    \mathcal{F}(\tilde{\mathcal{U}}, U) = \frac{1}{d^2(d + 1)} \sum\limits_{P \in \mathcal{P}_d} \trace{}{P^\dagger U^\dagger \tilde{\mathcal{U}}(P) U} +
    \frac{1}{d + 1},
\end{equation}
where $\mathcal{P}_d$ is the Pauli group defined on a $d$-dimensional Hilbert space \cite{SIgottesmanFaultTolerantQuantumComputation1999}. Since the device simulation is conducted over seven levels, but the RB experiment is performed over three levels, it is necessary to find the gate fidelity averaged over a subspace of states considered in the simulation, to obtain a gate fidelity comparable to the RB experiment.
To find the gate fidelity for the qutrit subspace, the elements of $\mathcal{P}_3$ are expanded into $7$-level operators by padding the matrices with zeroes such that $P \in \mathcal{P}_3$ occupies the top-left $3 \times 3$ block. $\mathcal{F}(\tilde{\mathcal{U}}, U)$ is then found using equation \ref{si:eqn:nielsen_formula} with $d = 3$. This is done because the elements of $\mathcal{P}_3$ are used as a basis for writing down the maximally-entangled qutrit state $\selfketbra{\phi} = \frac{1}{d^2} \sum\limits_{P \in \mathcal{P}_d} P^* \otimes P$, where $\ket{\phi} = \frac{1}{\sqrt{3}} (\ket{0}\ket{0} + \ket{1}\ket{1} + \ket{2}\ket{2})$. $\mathcal{F}(\tilde{\mathcal{U}}, U)$ is derived from the entanglement fidelity $F_e(U^\dagger \circ \tilde{\mathcal{U}}) = \braket{\phi | I_3 \otimes (U^\dagger \circ \tilde{\mathcal{U}}) | \phi}$, where $I_3 = \selfketbra{0} + \selfketbra{1} + \selfketbra{2}$, via 
\be
\mathcal{F}(\tilde{\mathcal{U}}, U) = \frac{d F_e(U^\dagger \circ \tilde{\mathcal{U}}) + 1}{d + 1},
\ee
following Ref.~\cite{SInielsenSimpleFormulaAverage2002a}. Since the qutrit maximally entangled state $\ket{\phi}$ has no components outside the lowest three levels, the elements of $\mathcal{P}_3$ cannot have support outside the space spanned by the lowest three levels, meaning $\mathcal{P}_3$ elements are embedded in larger Hilbert spaces by padding with zeroes. This also means $\mathcal{P}_3$ elements that are unitary in $3$ dimensions are not unitary in $7$ dimensions, since $P P^\dagger$ is not the $7$-dimensional identity operator $I_7$, but is instead $I_3$.

\begin{figure}
    \centering
    \includegraphics[width=6.8in]{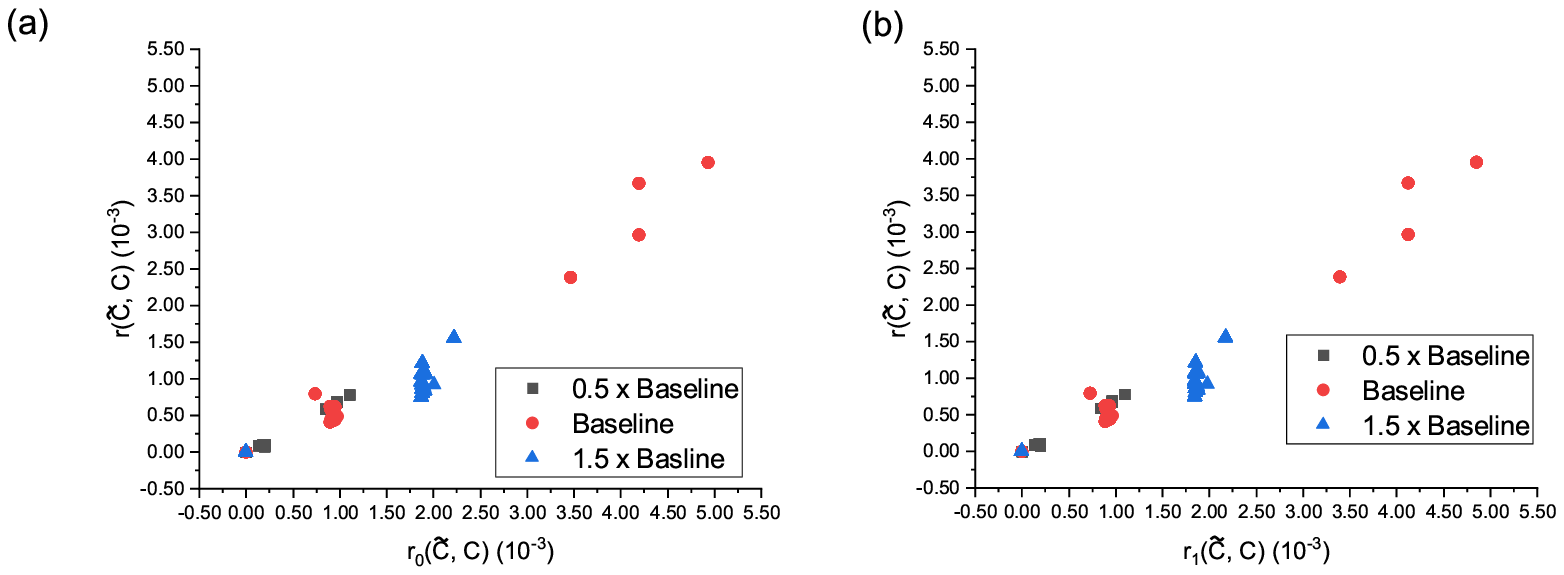}
    \caption{
    The simulated error for each Clifford gate versus its (a)
    zeroth-order estimate and (b) first-order estimate. The set labelled ``Baseline'' refers to a set calibrated using the experimental pulse amplitude. Multipliers refer to pulses where the amplitude was scaled proportionally and the pulse time was adjusted accordingly.}
    \label{si:fig:zeroth_and_first_approx}
\end{figure}

\section{Approximating the Fidelity of Clifford Gates from Givens Rotations}

Let $C$ represent an ideal Clifford gate and $\tilde{C}$ represent its error-prone implementation. $\tilde{C}$ is $\prod\limits_n \tilde{R}_n$, where $\tilde{R}_n = K_n R_n$ is a noisy implementation of an ideal Givens rotation $R_n$ with error $K_n$ where $n$ is a label for the constituent $R_n$ of a Clifford gate, ranging from $1$ to $3$. Assume $K_n = \alpha_n I + \beta_n M_n$, where $\alpha_n \in \mathbb{C}$, $\beta_n \in \mathbb{R}$, $M_n$ is an operator, and $I$ is the identity operator. The coefficient $\alpha_n$ is the complex number minimizing the magnitude of the largest eigenvalue of $I_3(K_n - \alpha_n I)I_3$, which is denoted by $\beta_n=||I_3(K_n - \alpha_n I)I_3||_\infty$. The three-level identity operator $I_3 = \selfketbra{0} + \selfketbra{1} + \selfketbra{2}$ and $M_n = \frac{1}{\beta_n} (K_n - \alpha_n I)$. The zeroth-order approximation to the fidelity $\mathcal{F}_0(\tilde{C}, C)$ assumes $\beta_n$ to be negligible. The fidelity expression then simplifies to
\begin{equation}
\begin{aligned}
    \mathcal{F}_0(\tilde{C}, C) &= \mathcal{F}\left(
    \prod_n K_n R_n, \prod_n R_n
    \right)\bigg|_{\beta_n = 0} \\
    &= \mathcal{F}\left(
    \prod_n \alpha_n \prod_n R_n, \prod_n R_n
    \right) \\
    &= \mathcal{F}\left(
    \prod_n \alpha_n I, I
    \right) \\
    &= \frac{1}{d^2(d + 1)} \sum\limits_{P \in \mathcal{P}_d} \trace{}{\left(
    	\prod_n \alpha_n P \prod_n \alpha_n^* P^\dagger
    	\right)} + \frac{1}{d + 1} \\
    &= \frac{1}{d^2 (d + 1)} \prod_n |\alpha_n|^2 \sum\limits_{P \in \mathcal{P}_d} \trace{}{P P^\dagger} + \frac{1}{d + 1} \\
    &= \frac{1}{d^2 (d + 1)} \prod_n |\alpha_n|^2 \sum\limits_{j = 1}^{d^2} \trace{}{I_d} + \frac{1}{d + 1} \\
    &= \frac{d \prod_n |\alpha_n|^2 + 1}{d + 1},
\end{aligned}
\end{equation}
where $I_d = \sum\limits_{n = 0}^d \selfketbra{n}$. Let $r_0(\tilde{C}, C) = 1 - \mathcal{F}_0(\tilde{C}, C)$ represent the zeroth-order approximation to the error. For a single Givens rotation $\tilde{R}_n$, $\alpha_n$ is related to $r_0(\tilde{R}_n, R_n)$ by
\be
|\alpha_n|^2 = 1 - \frac{d + 1}{d} r_0(\tilde{R}_n, R_n).
\ee
The zeroth-order error for a Clifford gate is then
\be
r_0(\tilde{C}, C) = \left(\frac{d}{d + 1}\right)\left(1 - \prod\limits_n \left(
1 - \left(\frac{d + 1}{d}\right) r_0(\tilde{R}_n, R_n)
\right)\right) .\label{si:eqn:zeroth_order_comp_mult}
\ee
Since $r_0(\tilde{R}_n, R_n)$ is small, one can make the approximation
\be
    \prod_n \left(
            1 - \left(\frac{d + 1}{d}\right) r_0(\tilde{R}_n, R_n)
    \right) \approx 1 - \left(\frac{d + 1}{d}\right) \sum\limits_n r_0(\tilde{R}_n, R_n),
    \label{si:eqn:mult_add_approx}
\ee
giving $r_0(\tilde{C}, C) \approx \sum\limits_n r_0(\tilde{R}_n, \tilde{R}_n)$. If $\beta$ is small, then $r(\tilde{C}, C) \approx \sum\limits_n r(\tilde{R}_n, R_n)$.

To get a more accurate approximation to $r(\tilde{C}, C)$, consider a gate $\tilde{C}$ made of two Givens rotations, such that $\tilde{C} = K_2 R_2 K_1 R_1$. To get the first-order approximation to the fidelity $\mathcal{F}_1(\tilde{C}, C)$, assume any terms with more than one $\beta$ are negligible when taking the product $P^\dagger C^\dagger \tilde{C} P \tilde{C}^\dagger C$. The approximation is given by
\be
    P^\dagger C^\dagger \tilde{C} P \tilde{C}^\dagger C \approx \alpha_2 \alpha_1 \alpha_1^* \alpha_2^*\left(
        I + P^\dagger R_1^\dagger R_2^\dagger \frac{\beta_2 M_2}{\alpha_2} R_2 R_1 P
        + P^\dagger R_1^\dagger \frac{\beta_1 M_1}{\alpha_1} R_1 P + I_d \frac{\beta_1 M_1^\dagger}{\alpha_1^*} + I_d R_1^\dagger \frac{\beta_2 M_2^\dagger}{\alpha_2^*} R_1
    \right). \label{si:eqn:first_order_ops_approx}
\ee
Substituting Eq.(\ref{si:eqn:first_order_ops_approx}) into Eq.(\ref{si:eqn:nielsen_formula}) gives
\be
\mathcal{F}_1(\tilde{C}, C) = \frac{d \prod\limits_n |\alpha_n|^2 + 1}{d + 1} + 
    \frac{\prod\limits_n |\alpha_n|^2}{d + 1} \left(
        \trace{}{\beta_1 \left(
            \frac{M_1}{\alpha_1} + \frac{M_1^\dagger}{\alpha_1^*}
        \right) I_d} + 
        \trace{}{\beta_2 \left(
            \frac{M_2}{\alpha_2} + \frac{M_2^\dagger}{\alpha_2^*}
        \right) I_d}
    \right), \label{si:eqn:first_order_f_two_rot}
\ee
which generalizes to
\be
\mathcal{F}_1(\tilde{C}, C) = \mathcal{F}_0(\tilde{C}, C) +
    \frac{\prod\limits_n |\alpha_n|^2}{d + 1} \sum\limits_n
        \trace{}{\beta_n \left(
            \frac{M_n}{\alpha_n} + \frac{M_n^\dagger}{\alpha_n^*}
        \right) I_d}. \label{si:eqn:first_order_f_n_rot}
\ee
Figure \ref{si:fig:zeroth_and_first_approx} shows $r(\tilde{C}, C)$ versus $r_0(\tilde{C}, C)$ and $r_1(\tilde{C}, C)$, for the gates in $\mathcal{C}_3$. The approximation improves as the gate error decreases. For these numerical results, decoherence was not included, so that the numerically-simulated propagator for each gate remains unitary, and only coherent control errors are considered.

\begin{figure}[!h]
    \centering
    \includegraphics[width=6.8in]{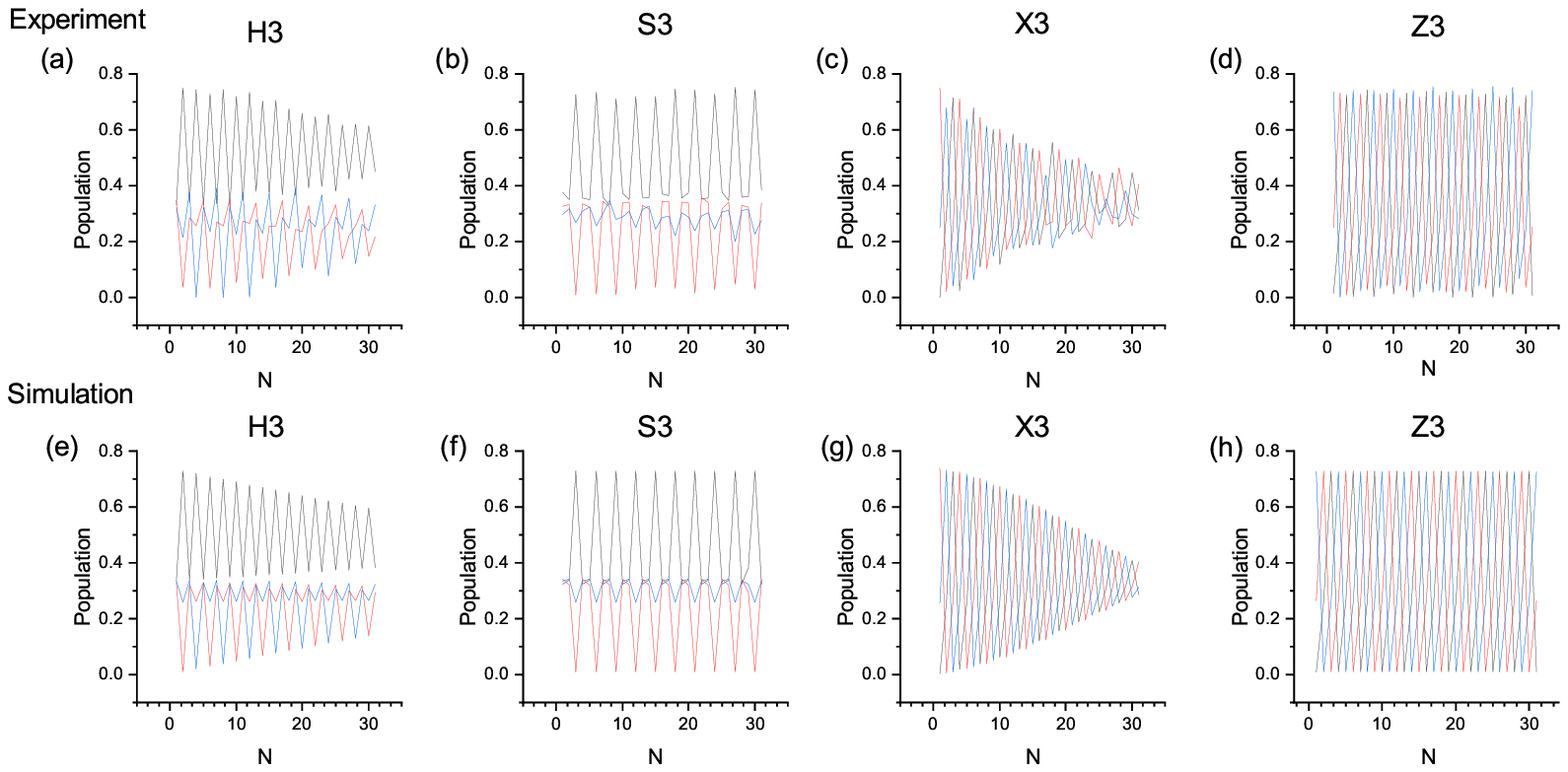}
    \caption{The experimentally-measured (top row, panels (a) to (d)) and simulated (bottom row, panels (e) to (h)) populations $P_0$ (black lines), $P_1$ (red lines), and $P_2$ (blue lines) versus the number of applications $N$ of each gate indicated.}
    \label{fig:ideal_and_simulated_pops}
\end{figure}

\section{Effective Hamiltonian Calculation}
The effective Hamiltonian $\heff$ for a given Givens rotation is found by minimizing the objective function $L = 1 - \mathcal{F}(\mathcal{U},U)$ over the space of possible Hamiltonians where $\mathcal{U}$ is the numerically simulated propagator for the rotation and $U=\exp({-i\heff \tau})$ with $\tau$ being the total duration for $\heff$. In the minimization function $L$, the fidelity $\mathcal{F}$ is calculated over seven dimensions instead of three in order for the effective Hamiltonian to include the dynamics outside the qutrit space as well. An ideal guess Hamiltonian $\tilde{H}$ is used as an initial condition for the minimization. The minimization is based on $\tilde{H}$ and $\heff$ being expressed as a weighted sum of Gell-Mann matrices $G_k$ with $\tilde{H} = \sum\limits_k \alpha_k G_k$ and $\heff = \sum\limits_k \beta_k G_k$ ~\cite{SIbertlmannBlochVectorsQudits2008}. During the minimization, $\beta_k$ is constrained to prevent spurious $2\pi$ rotations from being introduced.

\section{Simulated and Experimental Populations for Gate Repetitions}

Figure \ref{fig:ideal_and_simulated_pops} shows the experimental and simulated populations after $N$ repeated applications of the gates $H_3$, $S_3$, $X_3$, and $Z_3$. Simulations were conducted as described in Sec.~\ref{si:sec:numerical_simulations} including the measured decoherence rates given in Table~\ref{si:tbl:gamma_factors}.

\end{document}